\documentclass[lettersize,journal]{IEEEtran}
\usepackage{amsmath,amsfonts}
\usepackage{algorithmic}
\usepackage{algorithm}
\usepackage{array}
\usepackage[caption=false,font=normalsize,labelfont=sf,textfont=sf]{subfig}
\usepackage{textcomp}
\usepackage{stfloats}
\usepackage{url}
\usepackage{verbatim}
\usepackage{graphicx}
\usepackage{cite}

\usepackage{booktabs}   
\usepackage{tabularx}   
\usepackage{caption}    
\usepackage{makecell} 

\begin{document}

\title{A Robust Digital Receiver Architecture for Continuous-Variable Quantum Key Distribution}


\author{Lu~Fan$^{\dagger}$,
        Xuesong~Xu$^{\dagger}$,
        Miao~Yu$^{\dagger}$,
        Zhengyu~Li$^{*}$,
        Yunsong~Yin,
        Hao~Shi,
        Xuanyu~Guo,
        Song~Yu,
        and~Yichen~Zhang$^{*}$%
\thanks{$^{\dagger}$These authors contributed equally to this work.}%
\thanks{$^{*}$Corresponding authors: lizhengyu2@huawei.com; zhangyc@bupt.edu.cn.}%
\thanks{L. Fan, X. Xu, Y. Yin, S. Yu, and Y. Zhang are with the State Key Laboratory of Information Photonics and Optical Communications, School of Electronic Engineering, Beijing University of Posts and Telecommunications, Beijing 100876, China.}%
\thanks{Y. Zhang is with the Beijing Key Laboratory of Optical Quantum Information Network, Beijing University of Posts and Telecommunications, Beijing 100876, China.}%
\thanks{M. Yu is with the Department of Planning and Construction, China Mobile Communications Group Co., Ltd, Beijing 100032, China.}%
\thanks{Z. Li is with the Central Research Institute, 2012 Labs, Huawei Technologies Co., Ltd, Shenzhen 518129, Guangdong, China.}%
\thanks{H. Shi and X. Guo are with the Data Communication Research Department, Huawei Technologies Co., Ltd, Dongguan 523000, Guangdong, China.}%
}
\maketitle
\markboth{}%
\IEEEpubid{}

\maketitle

\begin{abstract}
Continuous-variable quantum key distribution utilizes advanced digital signal processing techniques to mitigate channel impairments, driving the progress of high-performance quantum secure communications.
Specifically, digital heterodyne receivers are commonly implemented using down-conversion algorithms to detect both components simultaneously where the signal is modulated onto an intermediate frequency to avoid low-frequency noise and bypass stringent optical phase-locking requirements.
However, a rigorous analysis of the quantum security implications of the digital down-conversion algorithm remains absent. 
In this paper, we investigate the algorithm process from the perspective of quantum operator evolution, revealing that the frequency folding effect inherent in down-conversion inevitably mixes the image-frequency information into the baseband signal. 
To address this issue, we propose a novel operator reconstruction detection scheme based on dual-path digital heterodyne receiver. 
By performing a non-equal weighted combination of the output operators along two optical paths in the digital domain, this scheme perfectly cancels out the image-frequency operator physically. 
The robust receiver architecture preserves the baseband noise-suppression advantages, inherently achieves perfect mode balancing for asymmetric optical detection losses, and fundamentally closes the image-frequency crosstalk.
Practical experiments have validated the feasibility of the proposed architecture and its effective suppression of image-frequency noise, achieving a performance of 11.4Mbps@42 km.
This work lays a solid theoretical and algorithmic foundation for future high-performance, highly secure digital continuous-variable quantum key distribution systems. 
\end{abstract}


\begin{figure*}[!h]
\centering
\includegraphics[width=12cm]{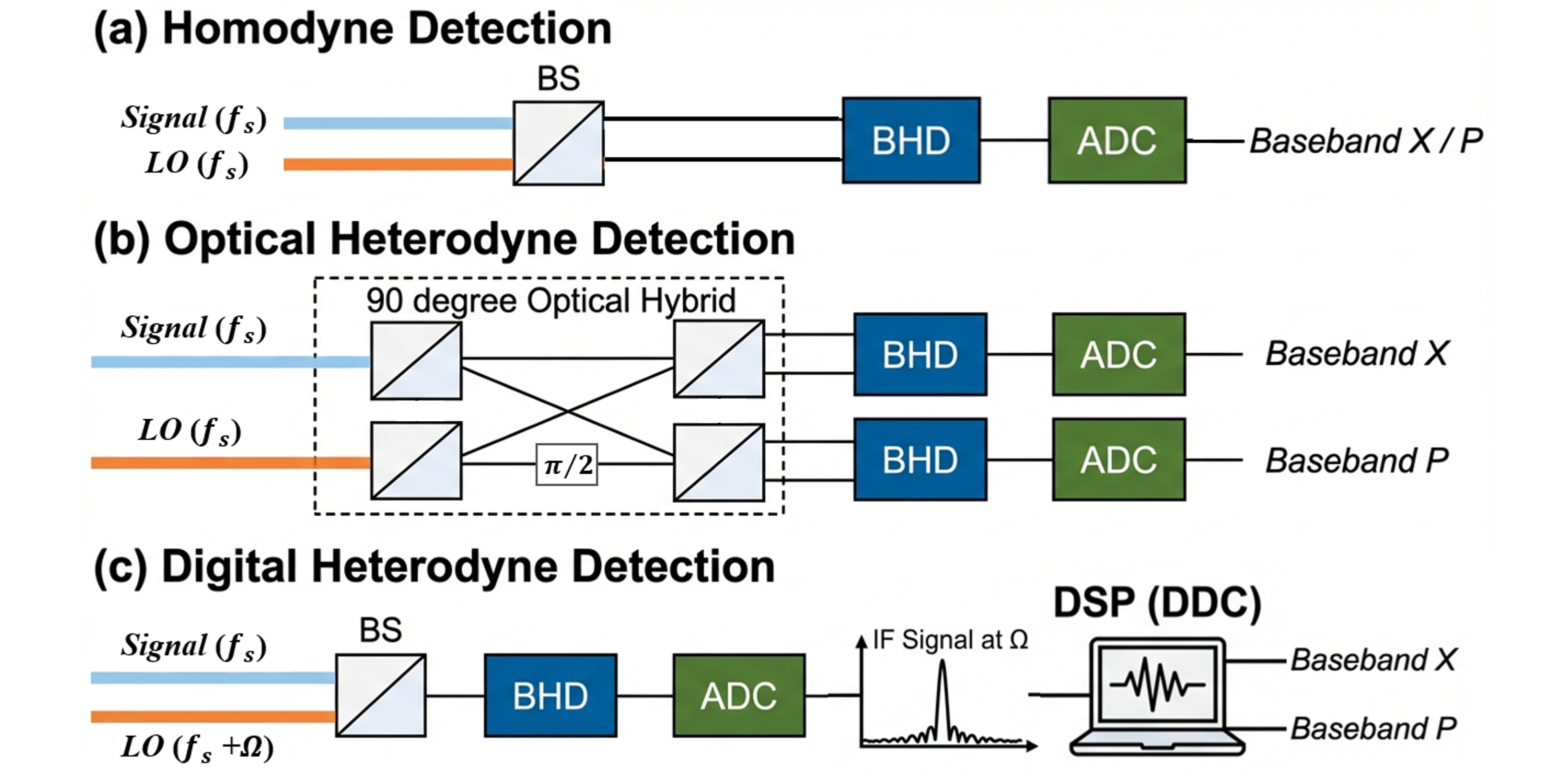}
\caption{The three current detection architectures for CV-QKD. (a) Homodyne detection: Signal and local oscillator (LO) at the same frequency are mixed via a beam splitter (BS) to measure a single baseband quadrature ($X$ or $P$) using one BHD. (b) Optical heterodyne detection: Signal and LO are mixed in a $90^\circ$ optical hybrid, simultaneously measuring $X$ and $P$ quadratures via dual BHDs. (c) Digital heterodyne detection: An LO with a frequency offset $\Omega$ generates an intermediate frequency signal, from which baseband $X$ and $P$ are simultaneously extracted via digital signal processing (DSP).}
\label{fig1}
\end{figure*}

\section{Introduction}
\IEEEPARstart{Q}
{uantum} key distribution (QKD) \cite{bennett2014quantum} provides a method for establishing secure cryptographic keys between two parties, with its security guaranteed by the fundamental principles of quantum mechanics.
Continuous-variable QKD (CV-QKD) \cite{grosshans2002continuous}  has emerged as a promising paradigm, which encodes key information into the quadrature components of the optical field \cite{pirandola2020advances,xu2020secure}. 
Benefiting from its high compatibility with standard commercial optical fiber communication \cite{zhang2024continuous,usenko2026continuous} and potential for high rates in metropolitan area, CV-QKD has attracted widespread attention and achieved remarkable progress in recent years. 
As systems evolve towards higher rates and longer distances \cite{jouguet2013experimental,zhang2019integrated,eriksson2019wavelength,zhang2020long,jain2022practical,tian2022experimental}, digital signal processing (DSP) technology is playing an increasingly central role in CV-QKD \cite{guo2021toward,jain2022practical,zhang2024continuous,  hajomer2024continuous1,roumestan2024shaped}. 
Digital CV-QKD systems not only efficiently compensate for device damages, channel drift, and phase noise but also drastically lower the rigorous demands on underlying optical hardwares.
Concurrently, the theoretical security foundations of DSP-based architectures are being progressively perfected, with increasing efforts devoted to bridging the gap between digital implementation and rigorous quantum security proofs \cite{qi2015generating,soh2015self,chen2023continuous,fan2026practical}.

In CV-QKD detection architectures, traditional homodyne detection requires strict frequency and phase locking between the signal and local oscillator (LO) \cite{jouguet2013experimental,qi2015generating,soh2015self}. 
Furthermore, the demodulated baseband signal is highly susceptible to low-frequency technical noises, such as laser relative intensity noise (RIN) and detector 1/f noise. 
To circumvent these baseband noise and eliminate the optical phase-locking constraint, the digital heterodyne detection architecture based on intermediate frequency (IF) modulation was introduced \cite{brunner2017low}. 
This architecture maintains a fixed frequency difference between the signal and the LO, capturing the quantum state containing both $x$ and $p$ information into an IF carrier using a single balanced homodyne detector (BHD). 
Subsequently, in the digital domain, the digital down-conversion (DDC) algorithm shifts the signal to the baseband and demultiplexes the two quadratures \cite{wang2022sub,hajomer2025experimental,wu2026high,zhang2026high}. 
This method physically reuses the hardware while functioning equivalently to optical heterodyne detection.

Although the DDC algorithm has demonstrated outstanding performance in modern CV-QKD experiments, a rigorous quantum security analysis of DDC algorithms for IF signals is still lacking.
In fact, the DDC process in the continuous-time domain can be described as the temporal mode filtering of quantum states at specific frequencies within the framework of quantum optics theory. 
At the signal frequency, this mode carries the quantum information modulated by the sender. 
However, due to the inherent real-valued impulse response of the band-pass filter (BPF) and its mixing with the digital carrier, the frequency folding effect inevitably aliases the image frequency component into the baseband which physically carries no information. 
In the standard entanglement-based (EB) model, this image frequency component should be equivalent to a vacuum state. 
Yet, in practical wideband optical receiving ends, it is always saturated with uncontrolled broadband technical noises (e.g., amplifier spontaneous emission noise). 
More severely, the existence of the image frequency provides a side-channel for an eavesdropper (Eve) who can manipulate the quantum state in the image band.

To address this security vulnerability whilst retaining the advantages of IF detection, this paper proposes a robust digital receiver architecture based on dual-path operator reconstruction.
Specifically, we upgrade the conventional single-BHD scheme to a true optical heterodyne setup utilizing dual-channel BHDs. 
Rather than processing the in-phase (I) and quadrature (Q) signals independently, we exploit the inherent phase correlation between the two paths during the digital down-conversion (DDC) process. 
By performing a precise, non-equal weighted coherent combination of the digitized I and Q signals, we reconstruct a new measurement operator, ensuring a highly effective coherent cancellation of the image-frequency modes from the two optical paths. 
Consequently, this architecture significantly suppresses the associated broadband excess noise, ideally achieving complete elimination under matched hardware responses.
Furthermore, its non-equal weighting mechanism naturally compensates for asymmetric optical detection losses between the I and Q paths, facilitating mode balancing without requiring strict hardware uniformity.
Experimentally, this architecture successfully mitigates the image-frequency interference, robustly suppressing the overall excess noise to a stable 0.034 shot noise unit, which consequently enables a high secret key rate of 11.4 Mbps over a 42 km standard single-mode fiber link.
By closing a critical security gap in digital receivers, this work paves the way for practical high-performance CV-QKD networks.

\section{Overview of the Digital Detection Architecture}
\subsection{Digital Receivers in CV-QKD.}
In CV-QKD systems, the key information is encoded in the orthogonal components of a coherent state, namely the in-phase component $X$ and the quadrature component $P$. 
As the quantum signal arriving at the receiver is extremely weak, suitable coherent detection techniques achieve noise-free amplification and demodulation of the weak quantum signal through beat generation with a high-power LO light.

The three current detection architectures for CV-QKD are shown in FIG. \ref{fig1}. 
The homodyne detection architecture in (a) is most commonly used in traditional CV-QKD systems which requires the central frequencies of the signal light field and the local oscillator light field to be exactly matched. 
After interference via a 50:50 beam splitter (BS) and detecting by BHD, the output baseband photocurrent is
\begin{equation}
    i(t)=2RA_{LO} [X(t)\cos(\Delta \theta(t))-P(t)\sin(\Delta \theta(t))]
\end{equation}
Here, $R$ denotes the responsivity of the photodetector, and $A_{LO}$ represents the complex amplitude of the LO light. $X(t)$ and $P(t)$ are the in-phase and quadrature components which carry quantum information, respectively. 
$\Delta\theta(t) =\theta_s (t)- \theta_{LO} (t)$ represents the instantaneous relative optical phase difference between the signal light and the LO light. 
Depending on the value of $\Delta\theta(t)$, $\Delta\theta(t) = 0$ corresponds to the $X$ component encoded by Alice, whilst $\Delta\theta(t) = -\pi/2$ corresponds to the $P$ component.

By orthogonally combining two BHDs, both components can be detected simultaneously. 
This constitutes the optical CV-QKD heterodyne detection scheme, as shown in FIG. \ref{fig1}(b). 
Although this detection architecture maximizes the utilization of the detection bandwidth, it still requires extremely precise optical locking between the signal and the LO. 
Furthermore, baseband detection is highly susceptible to interference from low-frequency technical noise, such as laser RIN and $1/f$ detector noise, resulting in a reduction in the system’s secret key rate.

To mitigate low-frequency baseband noise, the digital heterodyne detection scheme in FIG. \ref{fig1}(c) has been incorporated into the CV-QKD system. 
The optical frequency of the LO light maintains a fixed frequency difference $\Omega =\omega_{LO}-\omega_s $ with respect to the signal light. 
The output photocurrent is
\begin{equation}
    i_{IF}(t)=2RA_{LO} [X(t)\cos(\Omega t+\Delta \theta(t))-P(t)\sin(\Omega t+\Delta \theta(t))].
\end{equation}
In digital heterodyne detection, the two non-commuting quadrature components are simultaneously preserved. Physically, they are multiplexed onto two mutually orthogonal phases of the same IF carrier with $-\Omega$. Consequently, a single BHD is sufficient to acquire the complete information of the quantum state.
In the DSP stage, the DDC algorithm shifts the signal from the IF down to the baseband by mixing it with locally generated digital sine/cosine sequences, followed by low-pass filtering (LPF), as shown in FIG. \ref{DDC}(a). 
To address the residual phase drift $\Delta\theta(t)$, the system employs carrier phase estimation algorithms to perform phase compensation entirely within the digital domain, ultimately allowing for the precise orthogonal demultiplexing of the pure quadratures $X$ and $P$.

However, while the digital heterodyne architecture offers the advantages of eliminating low-frequency noise and restore both components simultaneously, it comes at the cost of an introduced image-frequency band. 
As conceptually illustrated in the spectral diagram of Fig. \ref{DDC}(b), the spectral convolution inherent in the DDC operation inevitably triggers a frequency folding effect in a single-BHD setup.
In the single BHD detection scheme, the spectral convolution inherent in the DDC operation inevitably triggers a frequency folding effect.  
This effect unavoidably aliases the uncontrolled broadband spectral components located at the image frequency $+\Omega$ directly into the desired baseband signal.
In the following section, we will provide an in-depth analysis of this image-band issue utilizing a rigorous quantum operator evolution framework.

\subsection{Quantum Operator Derivation and Security Analysis.}
\begin{figure}[t]
\centering
\includegraphics[width=8.5cm]{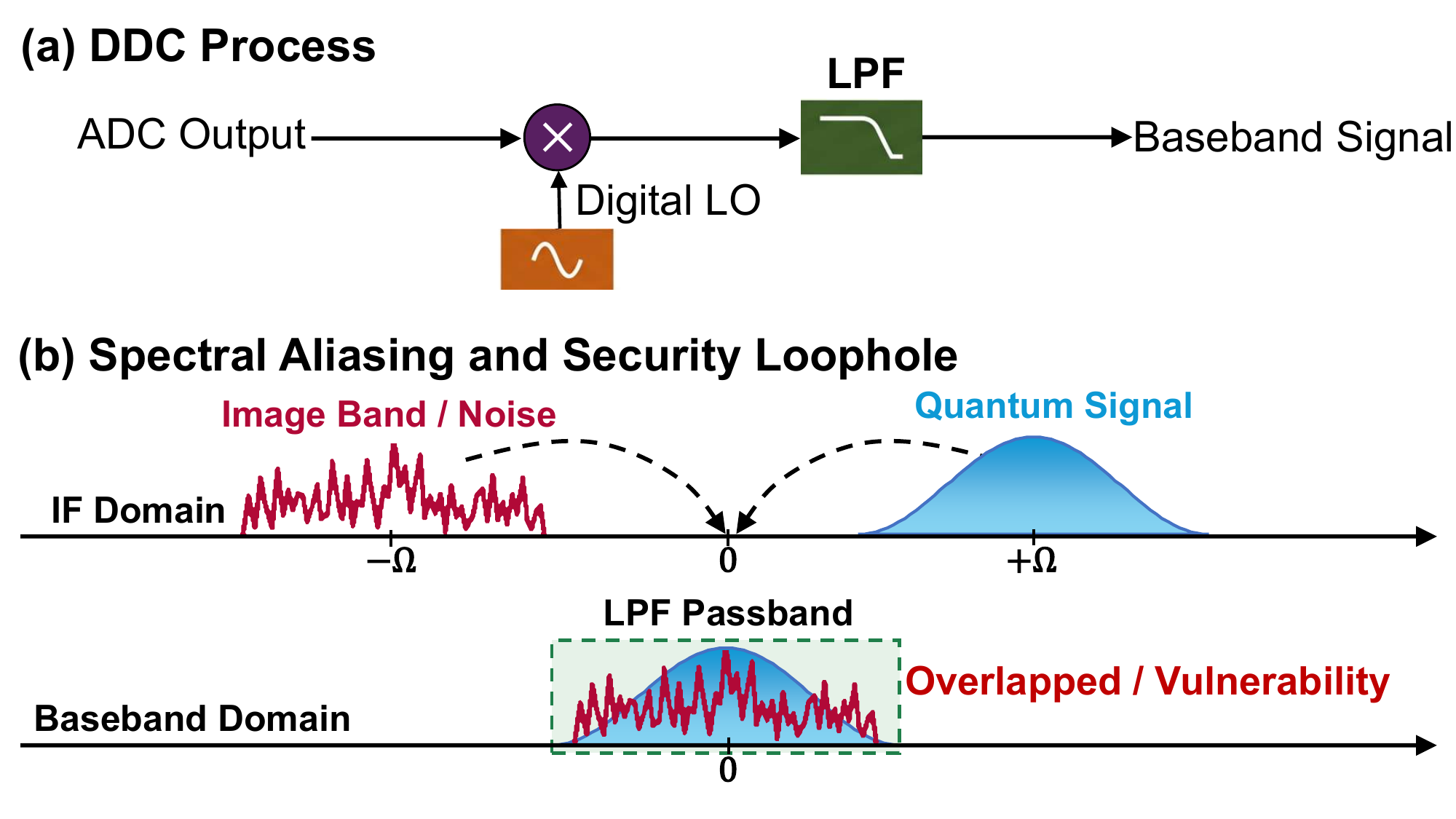}
\caption{(a) Diagram of DDC including down-conversion and filtering. 
(b) Frequency-domain spectral diagram demonstrating the frequency folding effect. During the digital mixing operation, spectral convolution inevitably aliases the uncontrolled broadband noise and potential side-channel manipulations from the image band directly into the desired baseband, contaminating the target quantum signal.}
\label{DDC}
\end{figure}

Based directly on the time-domain mode field theory, the instantaneous differential photocurrent operator after single-BHD is given by:
\begin{equation}
    \hat{f}^{LO}(t) = A_{LO} [\exp(-i\omega_{LO}t + i\theta_{LO}) \hat{a}^\dagger(t) + \exp(i\omega_{LO}t - i\theta_{LO}) \hat{a}(t)].
\end{equation}
Here, $\{\hat{a}^\dagger(t), \hat{a}(t)\}$ denote the time-domain operators of the quantum signal on a single BHD. 
Since the LO is a strong classical field, it can be expressed as ${a}_{LO}(t) = A_{LO} \exp(-i\omega_{LO}t + i\theta_{LO})$.

After passing through a detector with a finite bandwidth $h^{BPF}(t)$ and an amplifier, the operator becomes:
\begin{equation}
\begin{split}
    &\hat{f}^{BPF}(t)\\& = G_{LO} \int d\tau \left[ h^{BPF}(t - \tau) \exp(-i\omega_{LO}\tau + i\theta_{LO}) \hat{a}^\dagger(\tau) + h.c. \right],
    \end{split}
\end{equation}
where $ G_{LO}$ includes the gain of the coherent detection.

Following the ADC, the signal operator is
\begin{equation}
\begin{split}
    \hat{D}^{t_j}(t) = G_{LO} \frac{1}{\Delta t} \int d\tau \left[ \exp(-i\omega_{LO}\tau + i\theta_{LO}) \hat{a}^\dagger(\tau) +  h.c. \right] \\\int_{t_j}^{t_j+\Delta t} dt \, h^{BPF}(t - \tau).
\end{split}
\end{equation}

Then, the digital heterodyne algorithm necessitates multiplying the sampled results by $\cos(\Omega t + \varphi)$ in the digital domain. Along with the subsequent LPF summation, this operation can be mathematically approximated with high precision as a convolution in the continuous time domain. 
The output can thus be approximated as
\begin{equation}
    \hat{S}^{t_i} \approx G_{LO} \int d\tau \left[ \exp(-i\omega_{LO}\tau + i\theta_{LO}) \hat{a}^\dagger(\tau) + h.c. \right] h^{DDC}(t_i - \tau),
\end{equation}
where
\begin{equation}
\begin{split}
      h^{DDC}(t_i - \tau) = \frac{1}{2} e^{i\varphi} \left[ h^{BPF}(t_i - \tau) \exp(i\Omega t_i) \right] \ast h^{LPF}(t_i) \\+ \frac{1}{2} e^{-i\varphi} \left[ h^{BPF}(t_i - \tau) \exp(-i\Omega t_i) \right] \ast h^{LPF}(t_i)  
\end{split}
\end{equation}
represents the cascaded overall response of BPF, the multiplier, and LPF. 
The equivalent operation of the entire DDC link can be viewed as the original BPF response modulated by a complex exponential term, followed by a time convolution with the LPF.

\begin{figure*}[!h]
\centering
\includegraphics[width=14cm]{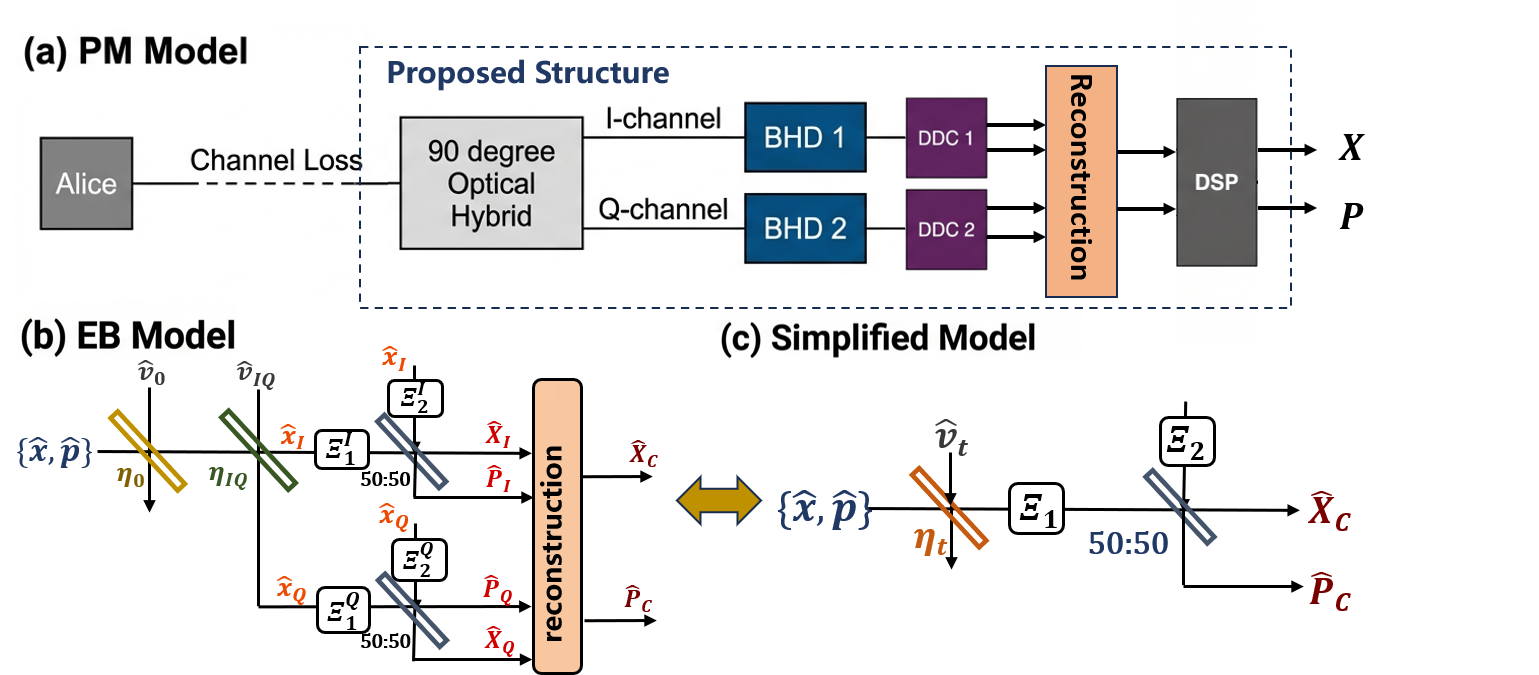}
\caption{(a) The PM model of proposed receiver structure. (b) The EB model of the proposed digital down-conversion with dual-BHDs receiver. 
Both the I-channel and Q-channel possess image frequency operator, enabling reconstruction to eliminate it.
(c) Further simplification of (b), where quantum state no longer contains image frequency operators.}
\label{dual}
\end{figure*}

Transforming this into the frequency domain provides further physical insight as shown in FIG.\ref{DDC} (b). The term $\exp(\pm i\Omega t_i)$ essentially performs spectral shifting. 
Due to the inherent real-valued impulse response characteristics of the BPF, its spectrum naturally contains symmetric negative and positive frequency components. 
During the DDC process, the quantum-frequency signal band originally located at $-\Omega$ and the image-frequency  band at $+\Omega$ are simultaneously folded and shifted to the baseband, both falling into the passband of the LPF to be extracted.

To rigorously formalize this frequency-folding mechanism within the time domain, we can decompose the overall DDC response $h^{DDC}(t_i - \tau)$ into two independent conjugate branches. 
Let us define $h^{DC-}(\tau)$ and $h^{DC+}(\tau)$ as the effective baseband envelope responses for the signal band and the image band, respectively. These effective envelopes represent the equivalent LPF results after the original BPF passbands centered at $-\Omega$ and $+\Omega$.

Based on this, we can equivalently devide the entire digital detection process into two independent quantum temporal mode wavepackets acting on the continuous optical field. 
The signal mode wavepacket associated with the negative-frequency band and the image mode wavepacket associated with the positive-frequency band can be written as
\begin{equation}
\begin{split}
        \xi_{1,-}(\tau) = h^{DC-}(\tau) \exp[-i(\omega_{LO} + \Omega)\tau], \\ 
        \xi_{2,+}(\tau) = h^{DC+}(\tau) \exp[i(\omega_{LO} + \Omega)\tau].
\end{split}
\end{equation}

Because the physical response of the filter is real-valued, the temporal mode of the image band is naturally the complex conjugate of the signal band's mode. 
By substituting these defined wavepackets back into the field operator, the output of the DDC process can be elegantly rewritten as:
\begin{align}
    \hat{S}^{t_i} = G_{LO} \int d\tau \Big[ &\exp(i\theta_{LO} + i\varphi) \hat{a}^\dagger(\tau) \xi_{1,-}(\tau) \nonumber \\
    + &\exp(-i\theta_{LO} - i\varphi) \hat{a}(\tau) \xi_{1,-}^*(\tau) \nonumber \\
    + &\exp(i\theta_{LO} - i\varphi) \hat{a}^\dagger(\tau) \xi_{2,+}(\tau) \nonumber \\
    + &\exp(-i\theta_{LO} + i\varphi) \hat{a}(\tau) \xi_{2,+}^*(\tau) \Big].
\end{align}

To interpret this output within the standard CV-QKD security framework, the field operators must be rigorously normalized corresponding SNU.
Extracting the scaling factor $G$ from the SNU, we redefine the unnormalized mode filtering into a pair of strictly normalized temporal modes $\Xi_{1,-}(\tau)$ and $\Xi_{2,+}(\tau)$.
As these two modes occupy distinct frequency bands, they are inherently orthogonal. This complete orthonormality is expressed as:
\begin{equation}
    \int \Xi_{i}^*(\tau) \Xi_{j}(\tau) \, \mathrm{d}\tau = \delta_{ij}, \quad \text{where } i, j \in \{(1,-), (2,+)\}
\end{equation}

By digitally selecting the DDC mixing phase $\varphi$, we can extract the corresponding quadrature components. 
Specifically, when $\varphi = 0$, the operation yields the in-phase quadrature:
    $X = \hat{x}_{\Xi_{1,-}} + \hat{x}_{\Xi_{2,+}}$ .
Conversely, setting $\varphi = -\frac{\pi}{2}$ extracts the quadrature phase:
    $P =  -\hat{p}_{\xi_{1,-}} + \hat{p}_{\xi_{2,+}} $.

In practical CV-QKD systems, the DSP operation at the transmitter only loads the Gaussian modulated signal onto the signal-frequency band $\Xi_{1,-}$, leaving the corresponding image band $\Xi_{2,+}$ without any data modulation. 
Therefore, in the EB theoretical model, the mode with wavepacket $\Xi_{2,+}$ entering the receiver can be essentially modeled as a quantum vacuum state. 
Up to this point, from the perspective of operator evolution, the single-path DDC algorithm shares the identical physical equivalent model with traditional optical heterodyne detection.

However, in practical broadband optical fiber communication links and photoelectric receiver front-ends, the image band is often permeated with various uncontrolled broadband technical noises. 
During the DDC multiplication process, these physical noises distributed within the image band are unconditionally aliased into the baseband along with the signal. 
Consequently, the receiver fails to purely extract the covariance matrix elements corresponding to the target signal operators $\{\hat{x}_{\xi_{1,-}}, \hat{p}_{\xi_{1,-}}\}$.
This not only dramatically increases the excess noise but also exposes an image side-channel that is highly susceptible to exploitation by Eve.

\section{A Operator Reconstruction Scheme Based on Dual-BHD Setup}

\subsection{The operator reconstruction method.}
To effectively eliminate the image-frequency vulnerability while retaining the advantages of IF modulation, we propose a robust operator reconstruction scheme based on dual-BHD optical heterodyne detection architecture. 

The overall conceptual framework of our proposed structure is illustrated in Fig. \ref{dual}. As depicted in the PM model in (a), the core intuition is to exploit the inherent phase relationship between the I and Q paths of a dual-BHD optical heterodyne front-end. 
By designing the digital phase of parallel DDC and performing weighted coherent combination, the reconstruction module systematically destructs the image-frequency modes in the digital domain. 
The EB model of the proposed architecture is shown in (b) and can be simplified to (c) following operator reconstruction, which definitively closes the image-frequency security loophole.

In the practical receiver, the $90^\circ$ optical hybrid and the dual BHDs inevitably exhibit slightly unbalanced quantum efficiencies, denoted as $\eta_1$ and $\eta_2$ for the I and Q paths, respectively. 
To elegantly map this hardware asymmetry into the operator evolution framework, we decouple the overall physical losses into a common-mode transmission $\eta_0 = \frac{1}{2}(\eta_1 + \eta_2)$ and a differential-mode splitting ratio $\eta_{IQ} = \frac{\eta_1}{\eta_1 + \eta_2}$. 

Mathematically, we treat the common-mode loss as an initial uniform attenuation, which transforms the incoming quantum signal $\hat{a}_0(t)$ into an intermediate operator $\hat{a}_1(t) = \sqrt{\eta_0}\hat{a}_0(t) + \sqrt{1-\eta_0}\hat{a}_{v_0}(t)$ and $v_0$ is the vacuum state. 

Subsequently, the differential-mode ratio $\eta_{IQ}$ acts as an beam splitter (BS) that divides this intermediate field into the I and Q paths. 
By incorporating the independent vacuum $\hat{a}_{v_{IQ}}(t)$ coupled through this asymmetric splitting, the resulting optical field operators are compactly expressed as:
\begin{equation}
\begin{aligned}
\hat{a}_I^\dagger(t) &= \sqrt{\eta_{IQ}} \hat{a}_1^\dagger(t) + \sqrt{1-\eta_{IQ}} \hat{a}_{v_{IQ}}^\dagger(t), \\
\hat{a}_Q^\dagger(t) &= -\sqrt{1-\eta_{IQ}} \hat{a}_1^\dagger(t) + \sqrt{\eta_{IQ}} \hat{a}_{v_{IQ}}^\dagger(t) ,
\end{aligned}
\end{equation}
where their Hermitian conjugates $\hat{a}_{I/Q}(t)$ follow analogously.

Based on the single-path DDC operator evolution derived previously, the down-converted output operators for paths I and Q can be written as
\begin{equation}
\begin{aligned}
&\hat{S}_I^{t_i} = \\&\sqrt{\eta_{IQ}} G^I \int d\tau \left[ \exp(i\theta_{LO} + i\varphi_I) \hat{a}_1^\dagger(t) \Xi_{1,-}^I(\tau) + h.c. \right. \\
&\left. + \exp(i\theta_{LO} - i\varphi_I) \hat{a}_1^\dagger(t) \Xi_{2,+}^I(\tau) + h.c. \right] \\
&+ \sqrt{1-\eta_{IQ}} G_{LO}^I \int d\tau \left[ \exp(i\theta_{LO} + i\varphi_I) \hat{a}_{v_1}^\dagger(t) \Xi_{1,-}^I(\tau) + h.c. \right. \\
&\left. + \exp(i\theta_{LO} - i\varphi_I) \hat{a}_{v_1}^\dagger(t) \Xi_{2,+}^I(\tau) + h.c. \right]
\end{aligned}
\end{equation}

\begin{equation}
\begin{aligned}
&\hat{S}_Q^{t_i}(t) = \\&-\sqrt{1-\eta_{IQ}} G^Q \int d\tau \left[ \exp(i\theta_{LO} + i\varphi_Q) \hat{a}_1^\dagger(t) \Xi_{1,-}^Q(\tau) + h.c. \right. \\
&\left. + \exp(i\theta_{LO} - i\varphi_Q) \hat{a}_1^\dagger(t) \Xi_{2,+}^Q(\tau) +h.c. \right] \\
&+ \sqrt{\eta_{IQ}} G^Q \int d\tau \left[ \exp(i\theta_{LO} + i\varphi_Q) \hat{a}_{v_1}^\dagger(t) \Xi_{1,-}^Q(\tau) + h.c. \right. \\
&\left. + \exp(i\theta_{LO} - i\varphi_Q) \hat{a}_{v_1}^\dagger(t) \Xi_{2,+}^Q(\tau) + h.c. \right]
\end{aligned}
\end{equation}

In the dual-BHD receiver setup, the $90^\circ$ optical hybrid establishes a rigid optical phase relationship between the LO of the I and Q paths: $\theta_{LO}^Q = \theta_{LO}^I + \frac{\pi}{2}$. To selectively eliminate the image band, we algorithmically lock the digital down-conversion phases such that $\varphi_Q = \varphi_I + \frac{\pi}{2}$. 

Assuming the S21 frequency responses of the balanced detectors are sufficiently matched (i.e., their temporal modes are identical, $\Xi_{1,-}^I(\tau) \approx \Xi_{1,-}^Q(\tau) = \Xi_{1,-}(\tau)$), we construct a newly recombined operator $\hat{S}_c$ through a precise weighted coherent combination of the I and Q outputs. Specifically, we cross-compensate for the differential-mode detection loss $\eta_{IQ}$ and the respective gains $G^I$ and $G^Q$ by defining the algorithmic reconstruction rule as
\begin{equation}
    \hat{S}_c = G^Q \sqrt{1-\eta_{IQ}} \hat{S}_I^{t_i} + G^I \sqrt{\eta_{IQ}} \hat{S}_Q^{t_i}.
    \label{eq:Sc_rule}
\end{equation}

By substituting the explicit forms of $\hat{S}_I^{t_i}$ and $\hat{S}_Q^{t_i}$ into this expression, the recombined operator is expanded as:
\begin{align}
    &\hat{S}_c\approx \\ &2G^I G^Q \sqrt{\eta_{IQ}(1-\eta_{IQ})} \int d\tau \left[\Xi_{1,-}(\tau) \exp(i\theta_{LO} + i\varphi_I) \hat{a}_1^\dagger(\tau) + \text{h.c.} \right] \nonumber \\
    &\quad + G^I G^Q (1 - 2\eta_{IQ}) \int d\tau \left[ \Xi_{1,-}(\tau)\exp(i\theta_{LO} + i\varphi_I) \hat{a}_{v_1}^\dagger(\tau)  + \text{h.c.} \right] \nonumber \\
    &\quad + G^I G^Q \int d\tau \left[  \Xi_{2,+}(\tau)\exp(i\theta_{LO} - i\varphi_I) \hat{a}_{v_1}^\dagger(\tau) + \text{h.c.} \right].
    \label{eq:Sc_expanded}
\end{align}

Crucially, as revealed in Eq.~(\ref{eq:Sc_expanded}), the quantum operator corresponding to the signal's image band ($\hat{a}_1^\dagger \Xi_{2,+}$) has perfectly vanished. 
Fundamentally, this cancellation is achieved by actively adjusting the phase in the digital domain. By deliberately aligning the digital DDC phase shift with the inherent optical phase difference of the hardware, we induce an exact $\pi$ phase inversion specifically for the image band and make it destructive interfere. 
As a result, the image-frequency components from the I and Q paths destructively interfere and vanish, while the target signal components constructively add up.
Meanwhile, the signal-frequency mode ($\hat{a}_1^\dagger \Xi_{1,-}$) constructively interferes, effectively doubling its coefficient.

To extract the normalized measurement quadratures, we calculate the SNU of this recombined operator as $\sigma_c^2 = 2\left|G^I G^Q\right|^2$. 
By normalizing $\hat{S}_c$ with respect to the $\sigma_c$, and setting the digital phase $\varphi_I = 0$ and $-\frac{\pi}{2}$, we obtain the reconstructed in-phase ($X_c$) and quadrature ($P_c$) components, respectively:
\begin{equation}
\begin{aligned}
    X_c &= \sqrt{2\eta_{IQ}(1-\eta_{IQ})} \hat{x}_{1,\xi_{1,-}} + \frac{1 - 2\eta_{IQ}}{\sqrt{2}} \hat{v}_{1, \xi_{1,-}} + \frac{1}{\sqrt{2}} \hat{v}_{2, \xi_{2,+}}, \\
    P_c &= -\sqrt{2\eta_{IQ}(1-\eta_{IQ})}\hat{p}_{1,\xi_{1,-}} - \frac{1 - 2\eta_{IQ}}{\sqrt{2}} \hat{v}_{1, \xi_{1,-}} + \frac{1}{\sqrt{2}} \hat{v}_{2, \xi_{2,+}}.
\end{aligned}
\label{eq:quadratures}
\end{equation}

\renewcommand{\tabularxcolumn}[1]{>{\centering\arraybackslash}m{#1}}
\begin{table*}[htbp] 
    \centering
    \caption{Comparison of technical characteristics and security aspects of different CV-QKD receiver architectures.}
    \label{tab:architecture_comparison_transposed}
    \renewcommand{\arraystretch}{1.5} 
    \begin{tabularx}{\textwidth}{@{} >{\raggedright\arraybackslash}X >{\centering\arraybackslash}X >{\centering\arraybackslash}X >{\centering\arraybackslash}X >{\centering\arraybackslash}X @{}}
        \toprule
        \textbf{\makecell{Receiver Architecture}} &
        \textbf{\makecell{Low-Frequency \\ Noise}} &
        \textbf{\makecell{Image-Frequency \\ Side-Channel}} &
        \textbf{\makecell{Asymmetric Loss \\ Robustness}} &
        \textbf{\makecell{Security Proof \\ Completeness}} \\
        \midrule
        Homodyne & Vulnerable (Baseband) & N/A & N/A & High \\
        
        Optical Heterodyne & Vulnerable (Baseband) & N/A & Poor (Hardware-reliant) & High \\
        Digital Heterodyne & Robust (IF band) & Vulnerable & Poor (Hardware-reliant) & Poor (Known loophole) \\
        
        \midrule 
        \textbf{Dual-Path Digital Heterodyne} &
        \textbf{Robust (IF band)} &
        \textbf{Robust (Physical Elimination)} &
        \textbf{High (Digital compensation)} &
        \textbf{High (Loophole closed)} \\
        
        \bottomrule
    \end{tabularx}
\end{table*}

Here, $\hat{v}_1$ represents the vacuum noise based on signal band, while $\hat{v}_2$ signifies the vacuum fluctuation entering from the unmodulated image band. 

Ultimately, since the $X_c$ and $P_c$ quadratures experience identical equivalent physical losses in this algorithm, we can factor out the intrinsic $3 \text{ dB}$ splitting penalty of traditional heterodyne detection. 
$\hat{x_0}$ and $\hat{p_0}$ in Equation \ref{eq:quadratures} are intermediate operators representing the evolution of original $x$ and $p$ via common-mode loss $\eta_0$.
Therefore, the overall equivalent detection efficiency $\eta_t$ for this symmetric, vulnerability-free model is remarkably given by:
\begin{equation}
    \eta_t = 4\eta_{0}\eta_{IQ}(1-\eta_{IQ}) = \frac{2\eta_1\eta_2}{\eta_1+\eta_2}.
\label{eq:efficiency}
\end{equation}
In this way, the scheme successfully achieves dynamic mode balancing while completely eliminating the image-frequency noise.

To contextualize the advantages of our proposed operator reconstruction scheme against existing technologies, Table \ref{tab:architecture_comparison_transposed} provides a comprehensive comparison of the key characteristics of several mainstream CV-QKD receiver architectures. 
As demonstrated by the preceding theoretical analysis, our dual-path digital heterodyne scheme, by physically eliminating the image-frequency band, not only fundamentally closes the security loophole inherent in digital heterodyne detection but also achieves high robustness against asymmetric loss via digital compensation. Consequently, it demonstrates significant superiority across all evaluated dimensions.

\begin{figure*}[!h]
\centering
\includegraphics[width=17cm]{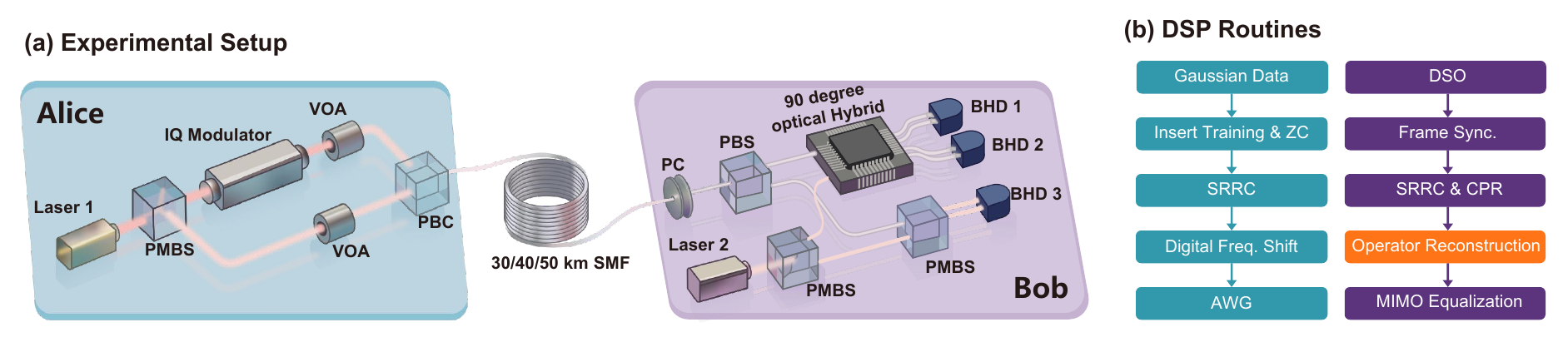}
\caption{(a) Experimental setup of the LLO CV-QKD system. IQ modulator, in-phase and quadrature modulator; PMBS, polarization
maintaining beam splitter; VOA, variable optical attenuator; PBC, polarization beam combiner; SMF, single-mode fiber; polarization controller.
(b) DSP routines. Crucially, the operator reconstruction scheme is applied to correct for hardware imbalances and suppressing image-band leakage after static DSP procedures. AWG, arbitrary waveform generator; ZC, Zadoff-Chu; SRRC, square root-raised cosine; DSO, digital storage oscilloscope; CPR, carrier phase recovery; MIMO, multi-input multi-output.
}
\label{exp_set}
\end{figure*}

\subsection{The residual operator leakage.}
In the ideal operator reconstruction presented earlier, the complete elimination of the image-frequency band fundamentally relies on the strict equivalence of the temporal modes between the I and Q channels. Importantly, since the weighted combination already mathematically normalizes the amplitude of both channels, any remaining discrepancy between the paths is purely a matter of pulse shape distortion (e.g., resulting from slightly mismatched balanced detector bandwidths or ADC responses). 
To formalize this imperfection, we first explicitly define the residual shape difference between the two paths' image-band wavepackets as $\Delta \Xi_{2,+}(\tau)$:
\begin{equation}
    \Delta \Xi_{2,+}(\tau) = \Xi^I_{2,+}(\tau) - \Xi^Q_{2,+}(\tau).
\end{equation}

By substituting this mismatch into the operator reconstruction process, the destructive interference is no longer perfect. 
The reconstructed output operator $\hat{S}_c$ will be contaminated by a residual leakage operator from the image band. By accurately retaining the physical coefficients derived in Eq.~(\ref{eq:Sc_expanded}), the actual measurement operator can be rigorously rewritten as
\begin{equation}
    \hat{S}_c \propto \underbrace{ \int \hat{a}_1^\dagger(\tau) [\Xi^I_{1,-}(\tau)+\Xi^Q_{1,-}(\tau)] \,d\tau}_{\text{Target Signal Band}} + \underbrace{\int \hat{a}_1^\dagger(\tau) \Delta \Xi_{2,+}(\tau) \,d\tau}_{\text{Residual Image-Band Leakage}}.
    \label{eq:Sc_leak}
\end{equation}



To precisely map this leakage to the standard SNU without any first-order approximations, we must accurately evaluate the variance of both the residual leakage band and the target signal band. Recognizing that the recombined signal mode mathematically evolves as the sum of the paths ($\Xi^I + \Xi^Q$), while the residual image mode evolves as the difference ($\Xi^I - \Xi^Q$), we can calculate their respective wavepacket integrals and the target signal band.
To quantify the excess noise coupled into the system, we evaluate the variance of the residual operator.
Given that the baseband wavepackets are inherently normalized ($\int |\Xi^I|^2 d\tau = \int |\Xi^Q|^2 d\tau = 1$), the variance integral naturally introduces the mode overlap factor $\gamma$, defined as $\gamma = \text{Re} \left( \int \Xi^I_{2,+}(\tau) \left[\Xi^Q_{2,+}(\tau)\right]^* \,d\tau \right)$ $(0 < \gamma \le 1)$. The integrals for the constructive (signal) and destructive (leakage) components are evaluated as:
\begin{align}
    \int \left| \Xi^I(\tau) + \Xi^Q(\tau) \right|^2 \,d\tau &= 2(1+\gamma), \nonumber \\
    \int \left| \Xi^I(\tau) - \Xi^Q(\tau) \right|^2 \,d\tau &= 2(1-\gamma).
\end{align}

To quantify the excess noise coupled into the system, we evaluate the variance of the residual operator. 
Given that the wavepackets $\Xi^I,\Xi^Q$ are normalized, expanding the integral of the residual wavepacket naturally introduces the mode overlap factor,

\begin{equation}
\begin{aligned}
    &\int \left| \Delta \Xi(\tau)_{2,+} \right|^2 \,d\tau \\&= \int \left| \Xi^I_{2,+} \right|^2 \,d\tau + \int \left| \Xi^Q_{2,+} \right|^2 \,d\tau - 2 \text{Re} \left( \int \Xi^I_{2,+}(\tau) \left[\Xi^Q_{2,+}(\tau)\right]^* \,d\tau \right) \\&= 2(1-\gamma),
\end{aligned}
\end{equation}
where $\gamma = \text{Re} \left( \int \Xi^I_{2,+}(\tau) \left[\Xi^Q_{2,+}(\tau)\right]^* \,d\tau \right)$ $(0 < \gamma \le 1)$ dictates the degree of mode matching between the two detection paths. 

Consequently, the relative mode-coupling weight ratio between the residual image band and the target signal band simplifies to
\begin{equation}
    \frac{W_{img}}{W_{sig}} = \frac{ \left( G^I G^Q \sqrt{\eta_{IQ}(1-\eta_{IQ})} \right)^2 \times 2(1-\gamma) }{ \left( G^I G^Q \sqrt{\eta_{IQ}(1-\eta_{IQ})} \right)^2 \times 2(1+\gamma) } = \frac{1-\gamma}{1+\gamma}.
\end{equation}

This derivation reveals a more profound physical security boundary. The ratio $W_\text{img} / W_\text{sig}$ described is, an energy transfer function that quantifies the coupling of the image-frequency band into the final signal measurement. 
It dictates that the energy of any attempt to influence the final measurement by manipulating the image band ($W_\text{img}$) will be strongly attenuated by a factor of $(1-\gamma)/(1+\gamma)$. 
This suppression is inherent to the receiver's physical mode matching $\gamma$ and is independent of Eve's specific strategy. 
For the common high degree of hardware matching, any energy leakage from the image band approaches zero, thus physically ensuring the purity of the target signal.

More compellingly, the operator reconstruction framework enables not only vulnerability elimination but also an active defense and monitoring mechanism. 
We can construct a monitoring operator, $\hat{S}_m$, which is mathematically orthogonal to the primary signal operator $\hat{S}_c$. 
By deliberately designing the reconstruction coefficients (e.g., by inverting the phase in the I/Q combination), this operator is engineered to have the opposite effect of $\hat{S}_c$: it actively cancels the contribution from the primary signal band while coherently enhancing the components from the image-frequency band.
By processing $\hat{S}_c$ and $\hat{S}_m$ in parallel within the digital domain, the receiver can simultaneously:
\begin{itemize}
    \item Robustly extract the secret key from $\hat{S}_c$, which is virtually immune to image-band interference.
    \item Detect any anomalous activity in the image band by monitoring the output power of $\hat{S}_m$ in real-time.
\end{itemize}
Any eavesdropping or attack targeting the image band would cause a significant increase in the energy of $\hat{S}_m$, providing the receiver with an unambiguous, real-time security alarm.

\section{Experimental Demonstration}
\subsection{Experimental Setup.}
The complete architecture of our CV-QKD system, encompassing both the optical hardware configuration and the DSP pipeline, is illustrated in Fig.\ref{exp_set}(a) and Fig.\ref{exp_set}(b).

As shown in Fig.\ref{exp_set} (a), at the transmitter side (Alice), a continuous-wave (CW) laser (Laser 1) operating in the telecom C-band is utilized as the optical source. The laser output is split into two distinct paths by a PMBS. The upper path is directed into an IQ modulator to encode the Gaussian-distributed quantum data, while the lower path bypasses modulation to serve as a strong, phase-correlated continuous pilot tone. To achieve the required quantum signal levels and establish an appropriate power ratio between the signal and the pilot, VOAs are applied independently to both paths. The quantum signal and the pilot tone are then orthogonally multiplexed in the polarization domain using a PBC and launched into the standard SMF.

At the receiver side (Bob), a PC and a PBS are first employed to demultiplex the incoming light, effectively separating the weak quantum signal from the strong pilot tone. A localized continuous-wave laser (Laser 2) acts as the LO. The LO is divided by a PMBS to serve both detection paths. The quantum signal is mixed with the LO using a $90^\circ$ optical hybrid and measured via BHD1 and BHD2, constituting a standard heterodyne detection scheme to acquire both the $X$ and $P$ quadratures. Concurrently, the pilot signal is mixed with the LO via a PMBS and measured by a third BHD3 in a homodyne configuration, which provides the necessary phase reference for mitigating laser phase noise.

Fig.\ref{exp_set} (b) depicts the comprehensive DSP workflow utilized in our system. The Tx DSP begins with the generation of true random Gaussian data. To facilitate robust channel equalization and synchronization at the receiver, training sequences (with an 8-times higher power level to ensure high signal-to-noise ratio) and ZC sequences are periodically inserted into the data stream. The composite digital signal undergoes upsampling and SRRC pulse shaping to constrain the signal bandwidth. Notably, a digital frequency shift is applied to up-convert the quantum data to an intermediate frequency of 1GHz, which effectively isolates the signal from low-frequency technical noise and baseband excess noise. The resulting digital sequences ($\{\tilde{X}_A, \tilde{P}_A\}$) are fed into an arbitrary waveform generator (AWG) operating at a high sampling rate to drive the optical IQ modulator.

At Bob's side, the electrical signals ($\{\hat{I}_B, \hat{Q}_B\}$) are captured by a high-speed digital sampling oscilloscope (DSO). 
The Rx DSP first utilizes the digital down-conversion model and simultaneously outputs the two captured electrical IQ signals.
The digital data undergoes ZC-based frame synchronization, downsampling, and matched SRRC filtering. The CPR is then executed by utilizing the phase information extracted from the pilot signal.
The proposed operator reconstruction scheme is applied here to suppress the image channel and balance the physical hardware, and outputs
reconstructed in-phase and quadrature components ($\{\hat{X}_c, \hat{P}_c\}$).
To normalize the quantum data, the SNU calibration data is captured with the quantum signal path blocked and processed through the exact same static algorithmic DSP flow.
Following this stage, to compensate for time-varying channel impairments and polarization crosstalk, the secure dynamic MIMO equalizer is trained using the extracted training sequences and applied to the quantum data. 

\subsection{Experimental Results.}
\begin{figure}[t]
\centering
\includegraphics[width=8cm]{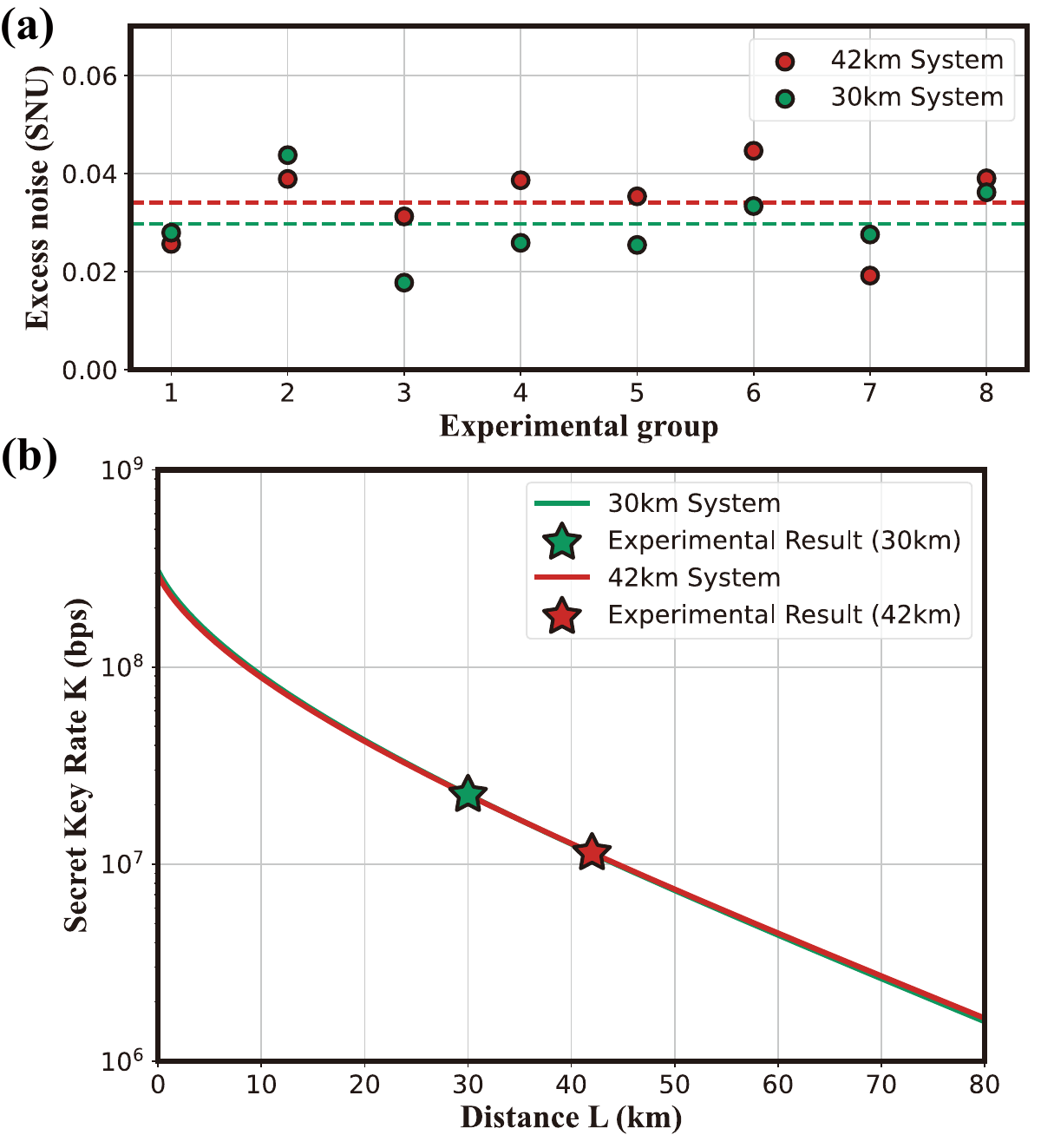}
\caption{(a) Estimated excess noise for eight groups of experimental data. The excess noise result of the green sphere is estimated under the 30km channel, with an average value of 0.029 SNU. Excess noise with 42km channel are
indicated by red spheres, with the average value underestimated to 0.034 SNU. 
(b) The secret key rate for 30km and 42km system with the proposed receiver structure. 
The solid line represents the theoretical
simulation curve based on the estimated system parameters.
The green and red star is general experimental secret key rate with 22.5Mbps@30km and 11.4Mbps@42km, respectively. 
}
\label{res}
\end{figure}

The performance of the proposed CV-QKD system and the secure digital down-conversion scheme was rigorously evaluated. In the experiment, the modulation variance $V_A$ of the quantum states at Alice's side was optimized and set to $V_A = 4$~SNU. 

The inherent detection impairments were modeled as optical losses. The experimental characterization revealed slightly unbalanced optical detection losses for the $I$ and $Q$ paths, measured at $\eta_I = 0.57$ and $\eta_Q = 0.51$, respectively. According to our theoretical model, the two regular components of the reconstructed output successfully share a unified and balanced equivalent detection loss of $\eta_t = 0.54$, verifying the effectiveness of the digital reconstruction module. 

For the channel parameter estimation, we collected 8 consecutive blocks of reconstructed data sets. Each data block contains approximately $N = 10 \times 10^6$ effective quantum symbols. 
Based on the covariance matrix of these blocks, the overall channel transmittance was estimated to be $T_c = 0.26, 0.163$ at $30$km and $42$km SMF, respectively. 
In Fig.\ref{res} (a), the 8 sets of excess noise estimated at $30$km and $42$km system with the proposed receiver structure
are shown as green and red spheres,
with mean values of 0.029 SNU and 0.034 SNU respectively.

The asymptotic secret key rate (SKR) is calculated utilizing the standard formula based on the Shannon mutual information ($I_{AB}$) between Alice and Bob, and the Holevo bound ($\chi_{BE}$) which represents the maximum information potentially leaked to Eve \cite{grosshans2004continuous}:
\begin{equation}
    K = f_{\text{repetition}} (1 - \alpha) (\beta I_{AB} - \chi_{BE})
\end{equation}
In this evaluation, the system repetition frequency is $f_{\text{repetition}} = 500$ MHz. The system overhead ratio, accounting for the insertion of training sequences, ZC sequences, and pilot tones, is fixed at $\alpha = 0.125$. Furthermore, a practical reverse reconciliation efficiency of $\beta = 0.96$ is assumed for the error correction phase.

The final experimental performance is presented in Fig.\ref{res} (b), which plots the secret key rate as a function of transmission distance. 
The solid line represents the theoretical simulation curve based on the estimated system parameters, while the green and red star denotes the actual experimentally extracted key rate at 30km and 40km system. 
The system successfully achieved an impressive secure key rate of 22.5Mbps@30km and 11.4Mbps@42km single-mode fiber link. The experimental data point aligns perfectly with the theoretical projection, firmly demonstrating the robustness, security, and practical feasibility of employing the secure digital down-conversion architecture in CV-QKD systems.

\section{Discussion and Conclusion}
In this paper, we have presented a comprehensive investigation into the quantum security vulnerabilities of digital IF detection architectures in CV-QKD. 
By employing a rigorous quantum operator evolution framework, we have identified that the frequency folding effect inherent in the DDC process inevitably aliases the unmodulated image-frequency mode into the baseband signal. 
This finding is critical because, in practical fiber-optic links, this image band is often saturated with uncontrolled technical noise or remains susceptible to side-channel manipulations by Eve, thereby compromising the security foundation of digital CV-QKD systems.

To address this challenge, we propose and experimentally validate a robust digital receiver architecture based on dual-path operator reconstruction. 
The core innovation lies in exploiting the inherent phase correlation between the I and Q paths of a dual-BHD setup to achieve physical cancellation of the image-frequency operators in the digital domain. The non-equal weighted coherent combinations naturelly provides an algorithmic pathway to compensate for hardware asymmetries, such as unbalanced quantum efficiencies of PDs and gain fluctuations. 
Our theoretical simulations and experimental data demonstrate that this reconstruction mechanism effectively relaxes the stringent requirements for hardware uniformity, achieving perfect mode balancing even with non-ideal optical components.
The experimental demonstration over a 42 km standard single-mode fiber link yielded a stable excess noise of 0.022 SNU and a high secret key rate of 11.4 Mbps. These results not only confirm the high-performance capabilities of the proposed architecture but also underscore its robustness against channel impairments. 

In conclusion, this work significantly provides a rigorous methodology to close a critical security gap in digital CV-QKD receivers through operator reconstruction. 
By fundamentally closing the side-channel vulnerabilities introduced by the DDC algorithm, this work ensures that digital receivers can maintain the same level of security integrity as traditional homodyne or heterodyne schemes while retaining the practical benefits of IF detection. 
By establishing a solid theoretical and algorithmic foundation, our research paves the way for the development of high-speed, robust, and physically secure quantum communication networks suitable for large-scale metropolitan deployment.

\section{Acknowledgments}
 This research was supported by the National Natural
Science Foundation of China (U24B20135), the National Cryptologic Science Fund of China (2025NCSF02050).
\bibliographystyle{IEEEtran}
\bibliography{references.bib}


\end{document}